\documentclass{ws-procs9x6}

\begin{document}

\title{Development of Acoustic Sensors for the {\sc Antares} Experiment
}

\author{C.~Naumann, G.~Anton, K.~Graf, J.~H{\"o}ssl, A.~Kappes, T.~Karg, U.~Katz, R.~Lahmann and
  K.~Salomon}

\address{Physikalisches Institut,\\ Friedrich-Alexander Universit{\"a}t Erlangen-N{\"u}rnberg,\\
  Erwin-Rommel-Stra{\ss}e 1,\\ 91058 Erlangen, Germany\\ E-mail: christopher.naumann@physik.uni-erlangen.de}

\maketitle 

\abstracts { In order to study the possibility of acoustic detection of ultra-high energy
  neutrinos in water, our group is planning to deploy and operate an array of acoustic
  sensors using the {\sc Antares} Neutrino telescope in the Mediterranean Sea. Therefore,
  acoustic sensor hardware has to be developed which is both capable of operation under the
  hostile conditions of the deep sea and at the same time provides the high sensitivity
  necessary to detect the weak pressure signals resulting from the neutrino's interaction
  in water. In this paper, two different approaches to building such sensors, as well as
  performance studies in the laboratory and {\it in situ}, are presented.
}

\section{Introduction}

At ultra-high energies, acoustic detection of cosmic neutrinos is a very promising approach, complementary
to optical detection in ice and sea water. Due to the much larger effective volumes, acoustic
detectors in water, ice or salt could reach far beyond the energies accessible to current detectors\cite{Karg1}.

Our group plans to integrate a number of customised detector units (storeys) for acoustic detection
into the {\sc Antares} detector, in which the optical sensor elements are replaced by ultrasound sensors. The aim
is to perform long term studies of the acoustic background in the deep sea and to investigate
the technical feasibility of acoustic particle detection in water.

\section{Sensor Development and Integration}

\begin{figure}[hb]
\begin{center}
\includegraphics[height=5cm]{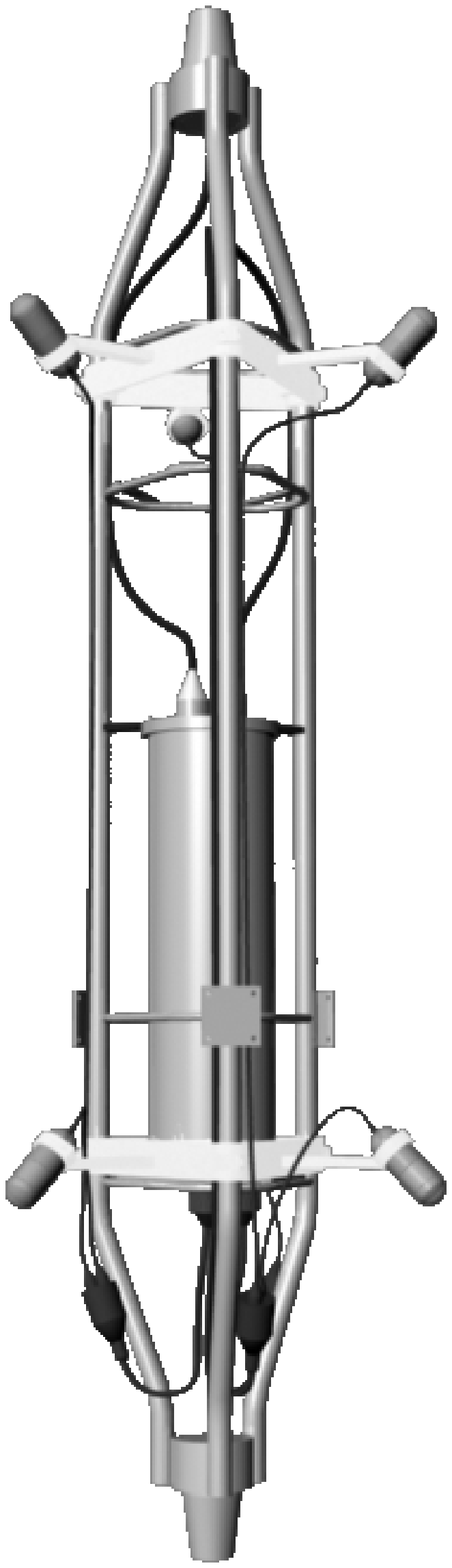}\hspace{4cm}
\nolinebreak
\includegraphics[height=5cm]{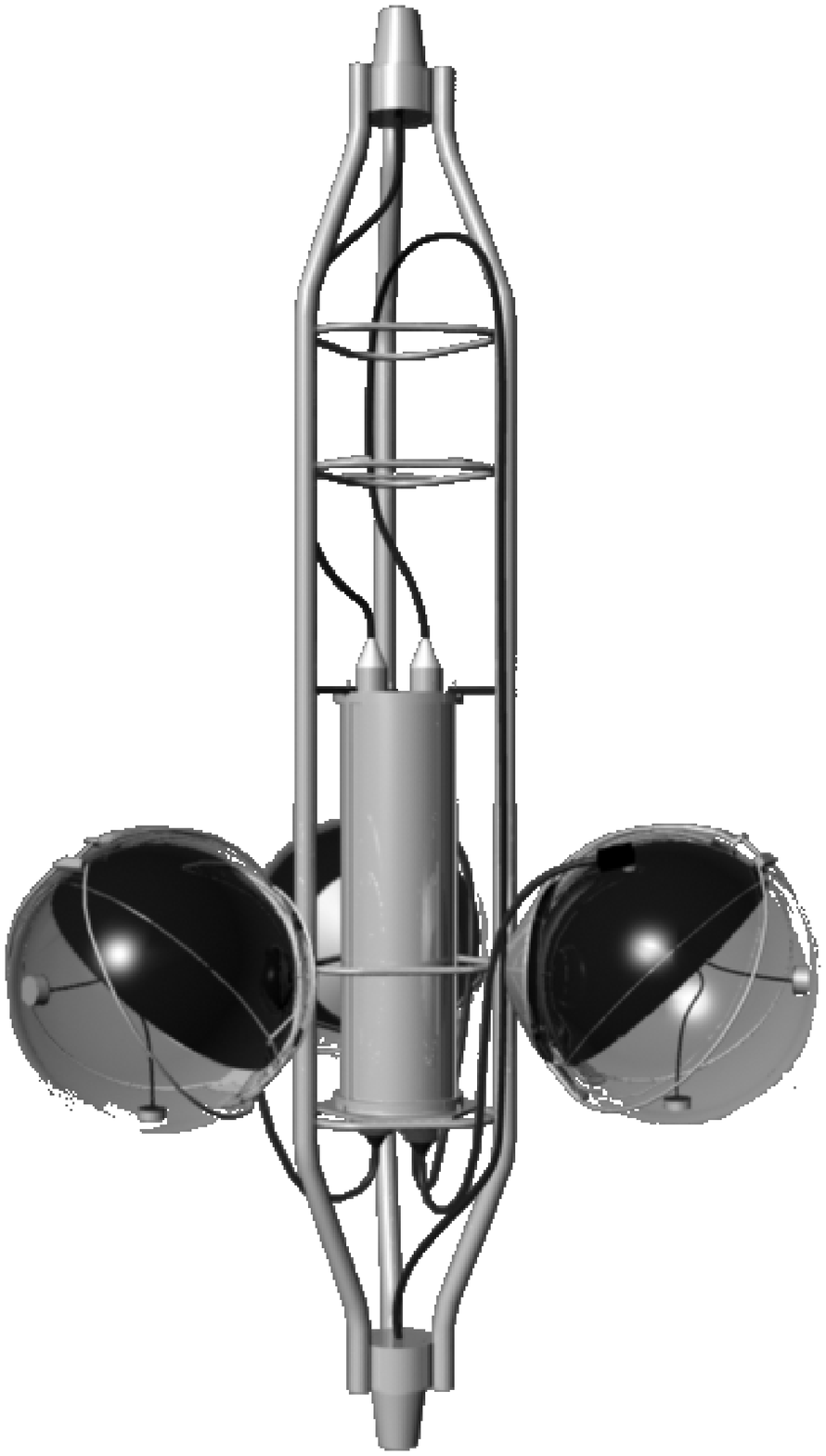}
\end{center}
\caption{Schematic design of an acoustic storey using hydrophones (left) and acoustic modules (right.)}\label{fig1}
\end{figure}

The sensors for a deep sea acoustic neutrino detector must be able to withstand the pressure
of several hundred bars and at the same time be sensitive to pressure
variations of the order of few mPa. For the integration into the {\sc Antares} detector,
it is also necessary to keep power consumption and changes to the overall detector design
as small as possible.
To meet these requirements, two different concepts have been proposed:
An array of individual hydrophones, replacing the optical modules (OMs)
of a ``standard'' storey, and modified glass spheres used as acoustic
modules (Fig.~\ref{fig1}). In both cases the data acquisition electronics is housed inside the {\sc Antares}
electronics cylinders.

The first concept uses an array of individual hydrophones, externally mounted on the support structure of an
{\sc Antares} storey with the OMs removed. As sensors, a mix of commercial and self-made hydrophones is forseen.
Commercial hydrophones, both used for civilian and military naval purposes, are not specifically
designed for the purpose of detecting neutrino signals with respect to their frequency response,
and their sensitivity is usually insufficient for neutrino detection. On the other hand, they
have the advantage of already providing the pressure- and water-resistance needed for deep sea operation. 
Therefore, a number of commercial hydrophones, as well as customised hydrophones bought from the
Institute for Theoretical and Experimental Physics, Moscow (ITEP), will be used.

In addition to the commercial hydrophones
, a range of different self-made hydrophones with piezoelectric tubes and/or discs as active
elements and custom-designed internal pre-amplifiers will be employed. 
As a protection against the sea water, both the piezo element and the pre-amplifier are
cast in polyurethane, chosen to match the surrounding water acoustically.
Both the sensitivity and the frequency characteristics depend on the size and shape of the piezo
elements, as well as the designs of the pre-amplifier, making it possible to build either
a broad-band sensor for background studies or one with its sensitivity optimised in a narrow
frequency band around the expected signal (about 5-50 kHz). The choice of geometry of the piezo
element also allows for different directional characteristics, varying from narrowly beamed to 
nearly $4\pi$ solid angle.

\vspace{13pt}
As a complementary approach, the spheres of the optical modules themselves are not removed 
but instead of the usual photomultiplier tubes fitted with acoustic sensor elements.
This makes the development of pressure- and water-resistant devices unnecessary as all equipment
is situated either inside the spheres or the electronics cylinders of the respective storeys.

As active sensor devices piezoelectric discs are used, several of which are attached
to the inside of each 17'' sphere, using polyurethane or epoxy resin as glue to ensure good acoustic
coupling to the glass. As in the case of the hydrophones, each piezo sensor is fitted with an
appropriate pre-amplifier directly mounted on the sensor to minimise noise pick-up.
The influence of the glass sphere on the signal shape and the overall sensitivity has yet to be
studied in detail.

\section{Sensitivity Studies and Sensor Calibration}

To quantitatively understand data from any acoustic sensor, its frequency dependence and directionality
must be known. Most of these studies can be done using a water tank in a laboratory.

As a sound source, a calibrated transducer is used, which is driven by short Gaussian voltage pulses from a signal generator.
With a digital oscilloscope, the signal response of the sensor is measured and compared to the
signal sent. From the Fourier transform of this transfer function, a relative calibration is
calculated, which only depends on the transducer characteristics. Using the transducer calibration
values provided by the manufacturer, an absolute sensitivity can be obtained.

\begin{figure}[hbt]
\begin{center}
\includegraphics[height=5.5cm]{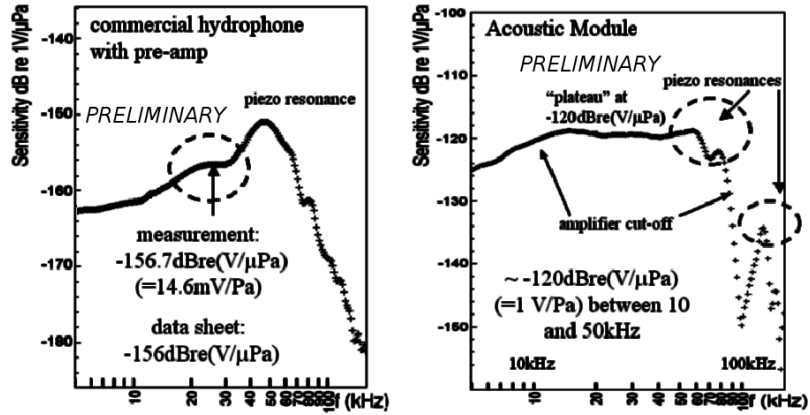}\label{fig2}
\end{center}
\caption{Frequency response of a commercial hydrophone (High Tech, Inc., left) and
an acoustic module prototype (right), both equipped with a pre-amplifier.
}
\end{figure}

Examples of a sensitivity spectrum for a commercial hydrophone and an acoustic module
prototype are given in fig.~\ref{fig2}. Below the first piezo resonance, the sensitivity is
nearly flat, giving a ``plateau'' at $-120dB\,re\;1V/\mu Pa$ for the sphere. This corresponds to
a signal of 1~mV per mPa, which should be sufficient for the expected pressure signals for neutrino
energies $\ge 10^{18} eV$ at a distance of 400~m. Both at low and high frequencies, the sensitivity
is cut off by the pre-amplifier to reduce the noise influence outside the signal region and
to avoid aliasing problems during digitisation.
The poorer sensitivity of the commercial hydrophone can be explained by the lower sensitivity
of the piezo element as well as the lower gain of its pre-amplifier.

\section{First {\it in situ} Measurements}

\begin{figure}[ht]
\begin{center}
\includegraphics[height=5.5cm]{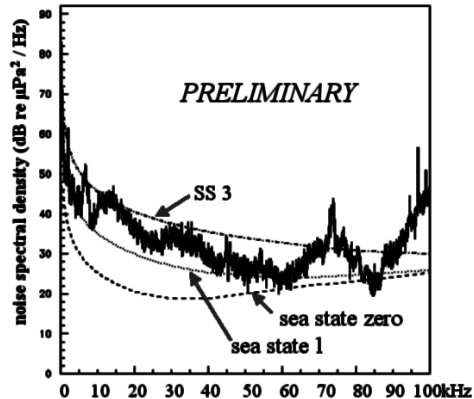}\label{fig3}
\end{center}
\caption{Preliminary noise spectrum at the {\sc Antares} site, measured with the {\sc Amadeus} detector,
  compared with predictions for sea state zero, one and three. The high-frequency excess can be explained by
  non-acoustic sources inside the container.}
\end{figure}

To test our sensor prototypes {\it in situ} and at the same time collect first environmental data
from the {\sc Antares} site,
an autonomous detector system ({\sc Amadeus}) was built from an {\sc Antares} electronics cylinder
fitted with piezo sensors and data acquisition hardware. Five piezo discs with pre-amplifiers were
glued to the inside of the titanium cylinder and read out via an autonomous PC board operated with
batteries. The whole setup was placed at the base of the {\sc Antares} test
line ``Line Zero'' and operated for several weeks at a depth of 2400~m, yielding about 15~GByte of data.
A full analysis of the data is still in progress but a first extraction of the environmental noise
spectrum (shown in Fig.~\ref{fig3}) is in good agreement with the expectations (derived from \cite{Urick1}).

\section{Conclusions}

For the installation of acoustic sensors into the {\sc Antares} detector, two different types
of acoustic detectors are planned, for both of which working prototypes have been built and
operated successfully. Data taken in the
laboratory and by an autonomous system in the deep sea show that the sensitivity needed for the
acoustic detection of UHE neutrinos can be achieved using custom-built piezoelectric sensors.

\section*{Acknowledgements}
This work is supported by the German BMBF Grant No. 05~CN2WE1/2.

\end{document}